# Fragment Isotope Distributions and the Isospin Dependent Equation of State

W.P. Tan[a], B-A. Li[b], R. Donangelo[c], C.K. Gelbke[a], M.-J. van Goethem[a], X.D. Liu[a], W.G. Lynch[a], S. Souza[c], M.B. Tsang[a], G. Verde[a], A. Wagner[a,1], H.S. Xu[a,2],

[a]*National Superconducting Cyclotron Laboratory and Department of Physics and Astronomy, Michigan State University, East Lansing, MI 48824, USA,* [b]*Department of Chemistry and Physics, Arkansas State University, State University, AR 72467, USA,* [c]*Instituto de Fisica, Universidade Federal do Rio de Janeiro, Cidade Universitaria, CP 68528, 21945-970 Rio de Janeiro, Brazil.*

## Abstract

Calculations predict a connection between the isotopic composition of particles emitted during an energetic nucleus-nucleus collision and the density dependence of the asymmetry term of the nuclear equation of state (EOS). This connection is investigated for central $^{112}$Sn+$^{112}$Sn and $^{124}$Sn+$^{124}$Sn collisions at E/A=50 MeV in the limit of an equilibrated freezeout condition. Comparisons between measured isotopic yield ratios and theoretical predictions in the equilibrium limit are used to assess the sensitivity to the density dependence of the asymmetry term of the EOS. This analysis suggests that such comparisons may provide an opportunity to constrain the asymmetry term of the EOS.

---

[1] Present address: Institut für Kern- und Hadronenphysik, Forschungszentrum Rossendorf, D-01314 Dresden, Germany
[2] On leave from the Institute of Modern Physics, Lanzhou, China

The equation of state (EOS) of strongly interacting matter governs the dynamics of dense matter in supernovae [1] and neutron stars [2,3]. Under laboratory-controlled conditions, the EOS has been investigated by colliding nuclei and measuring compression sensitive observables. The nuclear monopole and isoscalar dipole resonances, for example, sample the curvature of the EOS near the saturation density $\rho_0$ [4]. Measurements of the collective flow of particles emitted from the dense and compressed matter formed at relativistic incident energies can sample the EOS at densities as high as $4\rho_0$ [5]. In both types of experiment, investigations have primarily focused upon terms in the EOS that describe symmetric matter (equal numbers of protons and neutrons), leaving the asymmetry term that reflects the difference between neutron and proton densities largely unexplored [6]. For very asymmetric matter, however, details of this asymmetry term are critically important. For example, the asymmetry term dominates the pressure within neutron stars at densities of $\rho \leq 2\rho_0$, determines certain aspects of neutron star structure, and modifies proto-neutron star cooling rates [2,3].

Various studies have shown that the mean energy per nucleon $e(\rho,\delta)$ in nuclear matter at density $\rho$ and isospin asymmetry parameter $\delta=(\rho_n-\rho_p)/(\rho_n+\rho_p)$ can be approximated by a parabolic function

$$e(\rho,\delta)=e(\rho,0)+S(\rho)\delta^2 \qquad (1)$$

where $e(\rho,0)$ provides the EOS of symmetric matter, and $S(\rho)$ is the symmetry energy [2,3,6]. Different functional forms for $S(\rho)$ have been proposed [7], all consistent with constraints on $S(\rho_0)$ from nuclear mass measurements. Some theoretical studies have explored the influence of the density dependence of $S(\rho)$ on nuclear reaction dynamics [7-11].

Calculations of energetic nucleus-nucleus collisions [8-11] reveal that the relative emission of neutrons and protons during the early non-equilibrium stages has a robust sensitivity to the density dependence of $S(\rho)$. In general, pre-equilibrium neutron emission increases relative to pre-equilibrium proton emission for smaller values of the curvature $K_{sym}$ defined as:

$$K_{sym} = 9\rho_0^2 \left.\frac{\partial^2 S(\rho)}{\partial \rho^2}\right|_{\rho=\rho_0}. \qquad (2)$$

Enhanced pre-equilibrium neutron emission reduces the neutron-to-proton ratio in the dense region that remains behind [8,10].

Central collisions of complex nuclei of comparable mass provide the principal means to produce and study nuclear matter at densities either significantly above or below the saturation value. In near central Sn+Sn collisions at an incident energy of E/A=50 MeV, for example, matter is compressed to densities of about *1.5 $\rho_0$* before expanding and disassembling into 6-7 fragments with charges of *3≤Z≤30* plus assorted light particles [12]. Detailed analyses imply that such multifragment disassemblies occur at an overall density of *$\rho \approx \rho_0/6$-$\rho_0/3$* and over a time interval of about *$\tau \approx 30$-100 fm/c* [13-21]. Essentially all initial isotopic compositions are determined by the properties of the system during this narrow time frame when the density is significantly less than *$\rho_0$*. This implies that fragment isotopic distributions may have a significant sensitivity to the density dependence of S($\rho$). One can also enhance the sensitivity to the asymmetry term S($\rho$)·$\delta^2$ by varying the N/Z of the initial system.

Unfortunately, the observed isotopic distributions are also influenced by secondary decay, making it very important to identify observables that are insensitive to sequential decay. Statistical calculations have identified certain ratios of isotopic multiplicities as being robust with respect to the secondary decay [22,23]. For example, the ratio of the multiplicities $R_{21}(N_i, Z_i) = M_2(N_i, Z_i)/M_1(N_i, Z_i)$ of an isotope with neutron number *$N_i$* and proton number *$Z_i$* from two reactions 1 and 2 is relatively insensitive to the distortions from sequential decay. For multifragmentation, compound nuclear evaporation, and selected strongly damped collisions, such ratios as functions of *$N_i$* and *$Z_i$* have been experimentally shown to satisfy a power law relationship:

$$R_{21}(N_i, Z_i) = M_2(N_i, Z_i)/M_1(N_i, Z_i) = C\left(\hat{\rho}_p\right)^{Z_i} \left(\hat{\rho}_n\right)^{N_i}, \qquad (3)$$

where $\hat{\rho}_p$ and $\hat{\rho}_n$ are empirical parameters that have the interpretation, in the grand canonical approximation, of being the ratios of the free neutron and free proton densities in the two systems, $\hat{\rho}_p = \rho_{p2}/\rho_{p1}; \hat{\rho}_n = \rho_{n2}/\rho_{n1}$ [22]. One can also reduce the influence of secondary decay by taking ratios of the multiplicities of mirror nuclei *M($N_i,Z_i$)/ M($Z_i,N_i$)* measured in a single reaction [22,23], but the reduction of secondary decay effects may be less effective in this case.

The solid circles and squares in Fig. 1 show values for $\hat{\rho}_p$ and $\hat{\rho}_n$, respectively, obtained from fragments with $3 \leq Z_i \leq 8$ detected in central $^{112}$Sn+$^{112}$Sn, $^{112}$Sn+$^{124}$Sn and $^{124}$Sn+$^{124}$Sn collisions at E/A=50 MeV [22]. The $^{112}$Sn+$^{112}$Sn reaction was labeled as 1 in Eq. 3; the different data points correspond to the three choices for reaction 2 and are plotted in both left and right panels as a function of $N_{tot}/Z_{tot}$ where $N_{tot}$ and $Z_{tot}$ are the total numbers of neutrons and protons involved in reaction 2. The solid and open points in Fig. 2 show the experimental values for the mirror nuclei ratios constructed from the multiplicities of $^7$Li, $^7$Be, $^{11}$B and $^{11}$C fragments [22]. The upper and lower panels are for $^{124}$Sn+$^{124}$Sn and $^{112}$Sn+$^{112}$Sn collisions, respectively.

As discussed previously, the isospin asymmetries of the excited systems prior to multifragment breakup are sensitive to the density dependence of the asymmetry term of the EOS [8-10]. The "prefragment" is reduced in size relative to the total system by preequilibrium emission when it disintegrates into the final fragments. Both the Stochastic Mean Field (SMF) [24] and the Boltzmann-Uehling-Uhlenbeck (BUU) [25] formalisms, which describe the time evolution of the collision using a self-consistent mean field (with and without fluctuations, respectively), predict preequilibrium emission that is increasingly neutron-deficient and corresponding prefragments that are more neutron-rich for larger values of $K_{sym}$ [8,26]. These two formalisms are essentially identical during the early stages of the collision when the densities exceed $\rho_0/2$ and fluctuations in the mean field are negligible.

The mechanism for the disintegration of the prefragment into the observed fragments with $3 \leq Z \leq 30$ is an issue that is not settled but instead, is evolving considerably as new measurements and models become available. Dynamical multifragmentation models [14,27] have been used with some success, as have statistical models either with fragment emission probabilities determined from the rates for evaporative surface emission [28] or from the yields assuming thermal equilibrium [29,30]. Here, we examine the isotopic effects shown in Figs. 1 and 2 in the latter limit, which assumes that thermal equilibrium is achieved at breakup. Such calculations have provided surprisingly accurate predictions for the fragmentation of projectile- and target-like residues in peripheral and mid-impact parameter heavy ion collisions at incident energies $E_{beam}/A>200$ MeV [31,32], central heavy ion collisions at $E_{beam}/A \leq 50$ MeV [16,33] and in light ion induced collisions at $E_{beam} > 4$ GeV [34], after some accounting is made for preequilibrium light particle emission. Comparisons of experimental data to such approaches provide an assessment of the importance of non-equilibrium phenomena; accordingly, more difficulties in such approaches are encountered in central heavy ion collisions at $E_{beam}/A > 50$ MeV, reflecting the decreased time available for equilibration [33,35].

Specifically, we solved the BUU equation to obtain predictions for the dynamical emission of light particles during the compression and expansion stages of the collision. Then, we calculate the multifragment disintegration of the denser portions of the system via the Statistical Multifragmentation Model (SMM) of ref. [36,37]. In the first step of the hybrid calculations described here, the mean field for symmetric nuclear matter in the BUU calculations was chosen to have a stiff EOS (K=386 MeV) [38]. Calculations were performed with two different expressions for the asymmetry term, "asy-stiff" ($K_{sym}$=+61MeV) and "asy-soft" ($K_{sym}$=-69 MeV) [8,9]. Using these mean fields, BUU calculations were followed through the initial compression and subsequent expansion for an elapsed time of 100 fm/c at which point the central density decreased to a value of about $\rho_0/6$. The regions with densities $\rho>\rho_0/8$ were then isolated and their decay was calculated with the SMM.

The N/Z ratio and the nucleon number A of these fragmenting systems ("prefragments") are given in two leftmost columns in Table I. To illustrate the sensitivity of prefragment size and asymmetry to the elapsed time and density cutoff, values for N/Z and A are also given in Table I for an elapsed time of 80 fm/c. Calculations have shown that the N/Z ratio is not sensitive to the density cutoff [8]. While A is sensitive to these parameters, the N/Z ratio is relatively insensitive; to within 3%, values of N/Z of 1.27 (1.16), 1.36 (1.19) and 1.44 (1.23) are obtained for the source asymmetry of asy-stiff (asy-soft) calculations for $^{112}$Sn+$^{112}$Sn, $^{112}$Sn+$^{124}$Sn and $^{124}$Sn+$^{124}$Sn collisions independent of matching condition. The excitation energy per nucleon of the prefragment depends strongly on the matching condition; however, this quantity is presently difficult to calculate accurately. A range of values for the excitation energy per nucleon of $E^*/A$ = 4-6 MeV was therefore assumed in the subsequent SMM calculations to estimate the range of possible values consistent with the present approach.

Accurate calculations for isotopic yields from the multifragment decay of the excited prefragment within the SMM approach require a careful accounting of the structure and branching ratios of the excited fragments [23,36]. Using an SMM code [36,37] that carefully addresses such effects, the isotopic ratios in Figs. 1 and 2 were calculated for the prefragment source parameters in Table I. To indicate the sensitivity of these ratios to the secondary decay of heavier particle unstable nuclei, the open rectangles indicate the ratios obtained from the yields of primary fragments and the cross-hatched rectangles indicate the ratios obtained from the yields of the final fragments after secondary decay. The vertical height of each rectangle reflects the range of values for each quantity as the assumed excitation energy is varied over the range of $E^*/A$ = 4-6 MeV.

The left and right panels in Fig. 1 provide values calculated for prefragments obtained with the asy-stiff and asy-soft EOS's, respectively. In both panels, it can be seen that the ratios calculated from the primary yields (open rectangles) and those calculated from the secondary yields (cross-hatched rectangles) are similar, indicating that values for $R_{21}$ (N,Z) are relatively insensitive to secondary decay. With the exception of the value of $\langle \hat{\rho}_p \rangle$ for the $^{124}$Sn+$^{124}$Sn reaction, $N_{tot}/Z_{tot}$ =1.48, the ratios calculated from the final yields with the asy-stiff EOS (left panel) overlap the data. In comparison, the calculations using the asy-soft EOS (right panels) show a significantly weaker dependence on $N_{tot}/Z_{tot}$ than do the data.

The left and right panels in Fig. 2 provide values for the mirror nuclei ratios calculated with the asy-stiff and asy-soft EOS's, respectively. For these ratios, the sensitivity to the density dependence of the symmetry energy and to the secondary decay corrections are more significant. Ratios of mirror nuclei calculated with the asy-stiff EOS exceed those calculated with the asy-soft EOS by about a factor of two and overlap with the experimental values for three of the four ratios measured.

In the present simplified approach, the sensitivity of isotope and the mirror nuclei ratios to the asymmetry term arises from the different (N/Z) ratios of the prefragments that are predicted by BUU calculations. There is little sensitivity to the total mass of the prefragment, but additional sensitivity to its excitation energy per nucleon. Within the present model dependent analysis, this uncertainty in excitation energy is the limiting factor that prevents a more quantitative constraint on $S(\rho)$.

Light cluster emission during the early compression and expansion stages of the collision can influence the N/Z ratio and excitation energy of the prefragment. Incorporating the emission of light particles up to A=4 within transport model calculations will help address this issue [39,40]. While the present hybrid model approach demonstrates a sensitivity of the isotopic fragment yields to the asymmetry term of the EOS, the detailed nature of this sensitivity is model dependent. For example, the hybrid model predicts that an asy-stiff EOS leads to fragments that are more neutron-rich than those produced when the EOS is asy-soft. On the other hand, recent calculations with the Expanding Evaporating Source (EES) model, which assumes the fragments originate from surface emission and not from the equilibrium decay of the residue, predict the opposite trend [41]. It is therefore highly desirable to explore the connection between the fragment isotopic distributions and the EOS within other statistical and dynamical fragment production models currently in use and under development. These long-term goals require significant future theoretical efforts.

In summary, we have explored the connection between the isotopic composition of particles emitted during an energetic nucleus-nucleus collision and density dependence of the asymmetry term of the nuclear equation of state. This initial exploration was performed within the limit of an equilibrated freezeout condition. These calculations suggest that such data are sensitive to the density dependence of the asymmetry term of the equation of state. This work was supported in part by the National Science Foundation under Grant Nos. PHY-95-28844 and PHY-0088934, by Arkansas Science and Technology Authority Grant No. 00-B-14.by CNPq, FAPERJ, and FUJB and by the MCT/FINEP/CNPq (PRONEX) program, under contract #41.96.0886.00.

Table I: The first two columns provide the N/Z ratio and number of nucleons in the prefragments produced in the calculations for an elapsed time of 100 fm/c and density cutoff of $\rho_0/8$. The next two columns provide corresponding information for the same cutoff density but a shorter elapsed time of 80 fm/c. All calculations were performed at an impact parameter of 1 fm.

| reaction | t=100 fm/c, $\rho_c=\rho_0/8$ | | | | t=80 fm/c, $\rho_c=\rho_0/8$ | | | |
|---|---|---|---|---|---|---|---|---|
| | asy-soft | | asy-stiff | | asy-soft | | asy-stiff | |
| | N/Z | A | N/Z | A | N/Z | A | N/Z | A |
| $^{112}$Sn+$^{112}$Sn | 1.16 | 153 | 1.27 | 152 | 1.17 | 165 | 1.27 | 165 |
| $^{112}$Sn+$^{124}$Sn | 1.19 | 161 | 1.36 | 162 | 1.22 | 174 | 1.36 | 175 |
| $^{124}$Sn+$^{124}$Sn | 1.23 | 172 | 1.44 | 173 | 1.27 | 183 | 1.45 | 185 |

**Figure Captions:**

Figure 1: Both panels: The solid circles and solid squares in Fig. 1 show values for $\hat{\rho}_p$ and $\hat{\rho}_n$, respectively; measured in central $^{112}$Sn+$^{112}$Sn, $^{112}$Sn+$^{124}$Sn and $^{124}$Sn+$^{124}$Sn collisions at E/A=50 MeV. Left panel: the open and cross-hatched rectangles show corresponding hybrid calculations for $R_{21}$ calculated from the primary and final fragment yields, respectively, predicted by the hybrid calculations using the Asy-stiff EOS. Right panel: the open and cross-hatched rectangles show corresponding hybrid calculations for $R_{21}$ calculated from the primary and final fragment yields, respectively, predicted by the hybrid calculations using the Asy-soft EOS.

Figure 2: The solid and open points in the upper and lower panels show the mirror nuclei ratios measured for $^{124}$Sn+$^{124}$Sn and $^{112}$Sn+$^{112}$Sn collisions, respectively. Left panels: The open and cross-hatched rectangles show corresponding hybrid calculations of the mirror nuclei ratios calculated from the primary and final fragment yields, respectively, predicted by the hybrid calculations using the Asy-stiff EOS. Right panel: The open and cross-hatched rectangles show corresponding hybrid calculations of the mirror nuclei ratios calculated from the primary and final fragment yields, respectively, predicted by the hybrid calculations using the Asy-soft EOS.

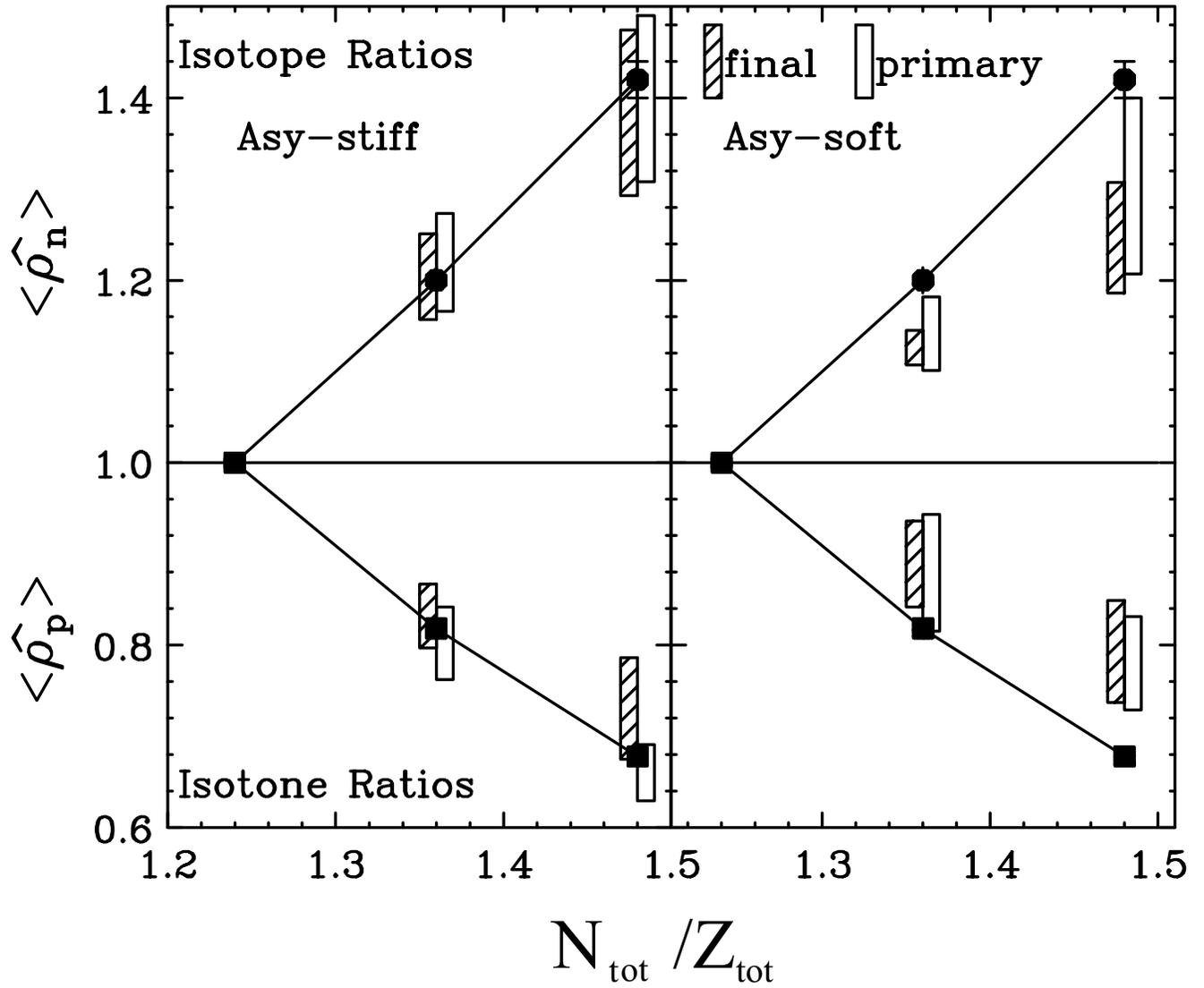

Fig. 1

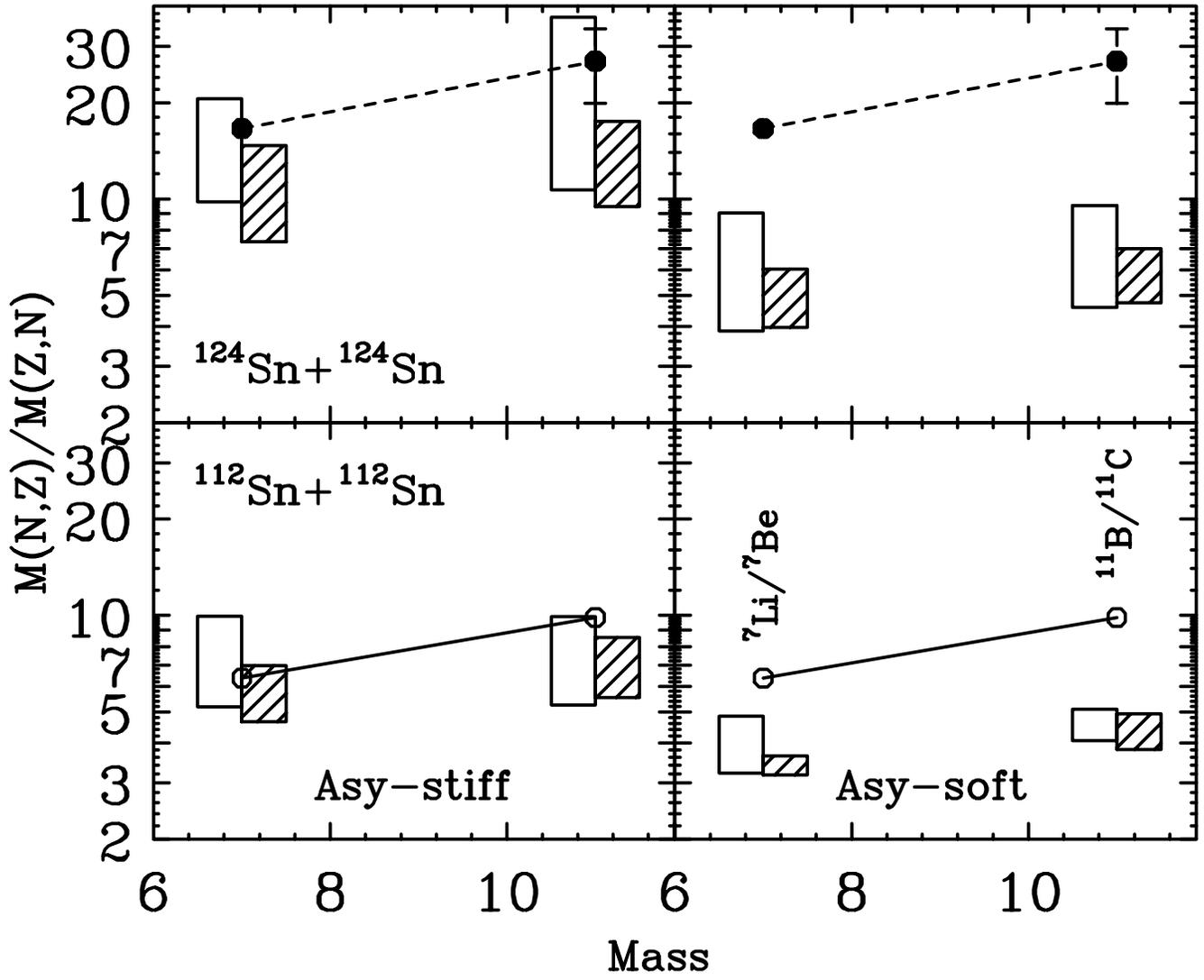

Fig. 2